\documentclass{article}
\usepackage{graphicx} 
\usepackage{siunitx}
\usepackage{listings}
\usepackage{hyperref}
\usepackage{nomencl} 
\makenomenclature
\usepackage{amsmath}
\usepackage{algorithm}
\usepackage{algpseudocode}
\usepackage{bm}
\usepackage{booktabs}
\usepackage[letterpaper, total={6in, 9in}]{geometry}
\usepackage{breqn}

\usepackage{multirow}
\usepackage{longtable}
\usepackage{booktabs}

\title{The Monte Carlo Computational Summit -- October 25 \& 26, 2023 -- Notre Dame, Indiana, USA
\footnote{This is an Accepted Manuscript of an article published by Taylor \& Francis in the Journal of Computational and Theoretical Transport on June 4 2024, available at: \url{https://doi.org/10.1080/23324309.2024.2354401}} \footnote{Please cite as: Morgan, J. P., Mote, A., Pasmann, S. L., Ridley, G., Palmer, T. S., \& Niemeyer, K. E. (2024). The Monte Carlo Computational Summit – October 25 \& 26, 2023 – Notre Dame, Indiana, USA. \textit{Journal of Computational and Theoretical Transport}, \textbf{53}(5), 361–382. \href{https://doi.org/10.1080/23324309.2024.2354401}{10.1080/23324309.2024.2354401}
}
\footnote{A full set of conference presentations available are at \href{https://doi.org/10.5281/zenodo.10207764}{zenodo.10207764}}}

\author{
    Joanna Piper Morgan$^{a,b}$\footnote{Contact: morgajoa@oregonstate.edu, joannapipermorgan@gmail.com},
    \and
    Alexander Mote$^{a,c}$
    \and
    Samuel Lee Pasmann$^{a,d}$
    \and
    Gavin Ridley$^{e}$
    \and
    Todd S. Palmer$^{a,f}$
    \and
    Kyle E. Niemeyer$^{a,b}$
    \and
    Ryan McClarren$^{a,d}$\footnote{Contact: mcclarr@nd.edu}
}

\date{%
    \small{
    $^{a}$Center for Exascale Monte Carlo Neutron Transport;  \\
    $^{b}$School of Mechanical, Industrial, and Manufacturing Engineering, Oregon State University;\\
    $^{c}$School of Electrical Engineering and Computer Science, Oregon State University;\\
    $^{d}$Department of Aerospace and Mechanical Engineering, University of Notre Dame;\\
    $^{e}$Department of Nuclear Science and Engineering, Massachusetts Institute of Technology; \\
    $^{f}$School of Nuclear Science and Engineering, Oregon State University
    }
}

\newcommand{\secsubtitle}[1]{\texorpdfstring{\\ \mdseries\itshape #1}{: #1}}

\begin{document}

\maketitle

\begin{abstract}
    The Monte Carlo Computational Summit was held on the campus of the University of Notre Dame in South Bend, Indiana, USA on 25--26 October 2023. The goals of the summit were to discuss algorithmic and software alterations required for successfully porting respective code bases to exascale-class computing hardware, compare software engineering techniques used by various code teams, and consider the adoption of industry-standard benchmark problems to better facilitate code-to-code performance comparisons.
    Participants reported that identifying and implementing suitable Monte Carlo algorithms for GPUs continues to be a sticking point. They also report significant difficulty porting existing algorithms between GPU APIs (specifically Nvidia CUDA to AMD ROCm).
    To better compare code-to-code performance, participants decided to design a C5G7-like benchmark problem with a defined figure of merit, with the expectation of adding more benchmarks in the future.
    The participants also identified the need to explore the intermediate and long-term future of the Monte Carlo neutron transport community and how best to modernize and contextualize Monte Carlo as a useful tool in modern industry.
    Overall the summit was considered to be a success by the organizers and participants, and the group shared a strong desire for future, potentially larger, Monte Carlo summits.
\end{abstract}


\section{Motivation}
The last decade has seen significant efforts to develop Monte Carlo neutral particle transport codes to better utilize accelerated compute platforms (i.e., GPUs)~\cite{Pereira2013, brantley2017llnl, Hamilton2019Continuous}, which represent the bulk of computational power in modern high-performance computers (HPC) and a significant amount in consumer-grade hardware ~\cite{Tramm2023}.
Developing algorithms and methods that can take advantage of GPUs can provide significant speedups to users, allowing for more rapid prototyping and analysis of nuclear systems at ever greater fidelity.

The parallelism paradigm on GPUs is considerably different from that on traditional CPUs,
meaning that algorithms that have been optimized for CPU-based performance have no guarantee of being equally efficient on GPUs.
In fact, this has been the experience with some code bases: initially seeing slow down when first porting to GPUs~\cite{pozzulp}.
This situation is only further complicated by an API environment that is far more fractured and disparate than the codified one seen on CPUs with shared memory parallelism from OpenMP and distributed memory parallelism from MPI.
Different vendors require the use of different proprietary toolchains that often do not translate well to other vendors' hardware.
Even support for seemingly basic operability on one vendor's system can not be guaranteed on other systems.
While there is support for open-source standards on multiple vendors' hardware (e.g., OpenCL), the functionality of these libraries often lags critically behind vendors' own APIs.
As a result, effectively porting to GPUs can require making major alterations to both underlying algorithms and to software engineering structures of a given code.
Different teams at different institutions have made different design choices for both basic algorithms and software engineering structures to achieve the same goal: computing ever more complex problems faster than ever.

As some codes have achieved the speedup promised by GPU-enabled stochastic transport others have had a more difficult time~\cite{pozzulp}. 
Each software and algorithmic design choice has its advantages and concessions; while some initially promising results have provided (at best) lackluster performance improvements, others have been shown to be invaluable to Monte Carlo performance on GPUs.
Finding a place where the Monte Carlo transport community could have frank discussions about what does and does not work was needed and provided the motivation for the Monte Carlo Computational Summit.

\section{Introduction}
The Monte Carlo Computational Summit was held on the campus of the University of Notre Dame in South Bend, Indiana on 25--26 October 2023. 
The goals of the summit were threefold: 
\begin{enumerate}
    \item Discuss algorithmic and software alterations required for successfully porting respective code bases to exascale-class computing hardware with heterogeneous architectures (i.e., systems with both CPUs and GPUs); 
    \item Compare software engineering techniques used by various code teams and significant roadblocks encountered; and
    \item Consider adopting industry-standard benchmark problems to better assess the comparative performance of various codes.
\end{enumerate}
The meeting was purposefully small, with 31 attendees from national laboratories and universities (listed in Table~\ref{tab:partcipants}) representing much of the stochastic neutral particle transport community in the United States.
The summit was organized by the Center for Exascale Monte Carlo Neutron Transport (CEMeNT) with the support of Alex Long (LANL) and Steven Hamilton (ORNL).

\begin{table}
    \centering
    {\begin{tabular}{@{}lc@{}}
        \toprule
        Institution & Number of Participants \\
        \midrule
        Argonne National Lab (ANL) & 4 \\
        Dept.\ of Energy/NNSA & 1 \\
        Lawrence Livermore National Lab (LLNL) & 5 \\
        Los Alamos National Lab (LANL) & 4 \\
        Massachusetts Institute of Technology (MIT) & 1 \\
        Naval Nuclear Lab (NNL) &  2\\
        North Carolina State University (NCSU)$^*$ & 2 \\
        Oak Ridge National Lab (ORNL) & 5 \\
        Oregon State University (OSU)$^*$ & 3 \\
        Sandia National Lab (SNL) & 1 \\
        Seattle University (SU)$^*$ & 1 \\
        University of Illinois; Urban-Champagne (UIUC) & 1 \\
        University of Notre Dame (ND)$^*$ & 2 \\
        \midrule
        Total & 31 \\
        \bottomrule
        
    \end{tabular}
    \caption{List of participating institutions in alphabetical order and the number of people attending from each ($^*$denotes CEMeNT university).}
    \label{tab:partcipants}
    }
\end{table}

\section{Presentations}
This section provides a brief overview and key highlights of the eight presentations given during the summit.
Table~\ref{tab:presentations} shows the title of each presentation along with the affiliated presenters, institutions, software, and ID listed in the presentation repository~\cite{mcclarren_2023_10207764}.
Table~\ref{tab:codes} shows the names of the codes discussed, their availability, distribution license (if any), and primary repository URL if publicly available.

\begin{table}
\centering
{\begin{tabular}{@{}lcccc@{}}
\toprule
Title & Presenter & Affiliation & Primary Code & ID \\ \midrule
Advances in the Shift GPU solver & Steven Hamilton & ORNL & Shift &  2\vspace{0.05in}\\ 
Implicit Monte Carlo at LANL--- & \multirow{2}{*}{Alex Long} & \multirow{2}{*}{LANL} & \multirow{2}{*}{Jayenne} & \multirow{2}{*}{3} \\
Progress and Challenges\vspace{0.05in}\\
MCATK GPU Prototype Code  & Tim Burke & LANL & MCATK &  N/A$^{\dagger}$\vspace{0.05in}\\
LLNL Monte Carlo Transport & Scott McKinley \& & \multirow{2}{*}{LLNL} & \multirow{2}{*}{Mercury} & \multirow{2}{*}{4} \\
GPU Status & Shawn Dawson & &\vspace{0.05in}\\
High Energy Physics Monte Carlo & \multirow{2}{*}{Seth Johnson} & \multirow{2}{*}{ORNL} & \multirow{2}{*}{Celeritas} & \multirow{2}{*}{6} \\
on GPU: \textit{Celeritas} & & & &\vspace{0.05in}\\
MC/DC+Harmonize: JIT Compilation & Ilham Variansyah \& & CEMeNT & \multirow{2}{*}{MC\slash DC} & 5a \& \\ 
\& Asynchronous GPU scheduling & Braxton Cuneo & (OSU\slash SU)&  & 5b\vspace{0.05in}\\
OpenMC at Exascale: Performance Portable & \multirow{3}{*}{John Tramm} & \multirow{3}{*}{ANL} & \multirow{3}{*}{OpenMC} & \multirow{3}{*}{N/A$^\dagger$} \\ 
Monte Carlo Particle Transport on \\ 
Intel, NVIDIA, and AMD GPUs & & & &\vspace{0.05in}\\
Experience Adapting OpenMC for CUDA$^*$ & Gavin Ridley & MIT & OpenMC & N/A$^\dagger$\vspace{0.05in}\\
Monte Carlo Research at NNL & Paul Burke & NNL & MC21 & 7 \\ 
\bottomrule
\end{tabular}%
\caption{$^*$Presenter absent, $^\dagger$presentations not made available for inclusion in the repository.}
\label{tab:presentations}
}
\end{table}

\begin{table}
    \centering
    {\begin{tabular}{@{}lcc@{}}
        \toprule
        Software name & Availability (license, if any) & Primary URL \\
        \midrule
        Shift & Distributed via RSICC & \url{rsicc.ornl.gov} \\
        Jayenne & Not publicly distributed & -- \\
        MCATK & Not publicly distributed & -- \\
        Murcury & Not publicly distributed & -- \\
        OpenMC CPU & Open source (specialized) & \url{github.com/openmc-dev/openmc} \\
        OpenMC GPU Argonne & Open source (MIT) & \url{github.com/exasmr/openmc} \\
        OpenMC GPU MIT & Open source (MIT) & \url{github.com/gridley/openmc/tree/cuda} \\
        MC21 & Not publicly distributed & -- \\
        MC/DC & Open source (BSD-3) & \url{github.com/CEMeNT-PSAAP/MCDC} \\
        Harmonize & Open source (BSD-3) & \url{github.com/CEMeNT-PSAAP/harmonize}\\
        Celeritas & Open source (Apache 2.0) & \url{github.com/celeritas-project/celeritas} \\
        \bottomrule
    \end{tabular}
    \caption{Software presented at the Monte Carlo Summit, their availability, licenses, and primary URLs.}
    \label{tab:codes}
    }
\end{table}

\subsection{Advances in the Shift GPU Solver \secsubtitle{Steven Hamilton, Oak Ridge National Laboratory}}
Shift is a continuous-energy CPU and GPU transport code, maintained at Oak Ridge National Laboratory \cite{PandyaTaraM.2016Icab}. 
Shift was originally written as a history-based CPU code with a rich set of features for modeling entire reactor systems. 
Many of these features have been ported to a separate GPU event-based code \cite{HamiltonStevenP.2021Ddit}. 
Within the event-based scheme, the CPU essentially acts as an event dispatcher, calling events and passing the relevant data to the GPU.
This means that the GPU and CPU back-ends share much of the same code, save for some data structure optimization and particle history re-implementation. 

Thanks to the minimal code alterations required for GPU implementation, the Shift development team has had success porting many CPU features to the GPU code including:
\begin{itemize}
    \item On-the-fly Doppler broadening,
    \item Thermal resonance upscattering,
    \item Domain decomposition,
    \item Isotopic depletion,
    \item Hybrid deterministic methods, and
    \item General geometry support.
\end{itemize}

Shift alternates between calling CUDA and the Heterogeneous-Compute Interface for Portability (HIP) for compiling to Nvidia and AMD GPUs, respectively.
HIP supports compilation on both Nvidia and AMD GPUs; however, in practice the ability to compile for Nvidia devices seems to be broken more often than it is functional.
To avoid splitting up the code base between CUDA and ROCm APIs, the Shift development team uses a Cmake macro to to replace relevant function calls.
The Shift team has identified that most (but not all) algorithmic optimizations for one GPU vendor will benefit the other, but AMD GPUs seem to be more sensitive to register usage and occupancy.
The number of registers that the compiler determines are needed by compute kernels is widely different depending on ROCm versions. 
This incurs random performance variations between versions of the ROCm compiler stack, meaning there is no guarantee that newer compiler versions will bring better performance, and in fact may result in worse runtimes without careful optimization.
The Shift team reports that every time they get a new ROCmm version they have to re-evaluate register use on a kernel-by-kernel basis, taking up valuable development time.
Overall, much more time is spent porting the build than the code itself.
Performance comparisons revealed that one Graphics Compute Die (GCD) of an AMD MI250X (which has two GCDs in total) was about equal to the performance of one Tesla V100, which in turn equates to about 150--200 CPU cores.

Going forward, the Shift development team will continue to make optimizations for the existing event-based GPU code and will continue to port more features to their GPU version.

\subsection{Implicit Monte Carlo at LANL---Progress and Challenges \secsubtitle{Alex Long, Los Alamos National Laboratory}}
Jayenne is a C++ implementation of the Implicit Monte Carlo (IMC) method described by~\cite{Fleck1971} to solve thermal radiation transfer problems. Jayenne is called by the user using a Fortran API as part of a larger hydrodynamics code package \cite{ThompsonKellyGlen2021JPMR}.
The hydrodynamic ``host'' code passes in the relevant temperatures and opacities.
Jayenne was first written at Los Alamos over 25 years ago, where it is still actively maintained and developed.
This presentation covered the recent effort to port Jayenne from CPUs to GPUs, which required major refactoring. 

The code was moved away from object-oriented design patterns and towards a functional programming design.
Several additional changes were made, including removal of virtual and recursive functions and alterations to the ``contractor'' class which contained many shared pointers---causing significant slowdown.
The Jayenne team focused on porting the most-used features first.
Other algorithmic and software optimizations implemented include:
\begin{itemize}
    \item Simplifying particle data structure to 128 bytes and cell tallies to be only 88 bytes;
    \item Allocating one particle per thread with particles immediately loaded into shared memory;
    \item Removing all virtual and recursive functions;
    \item Implementing various algorithm acceleration techniques after an un-accelerated version (e.g., Discrete Diffusion Monte Carlo (DDMC), Random Walk acceleration); and
    \item Using Cmake macros for to avoid splitting a code base between Nvidia and ROCm versions.   
\end{itemize}

These optimizations resulted in 9$\times$ speedups in IMC on smaller transport-dominated problems and 6$\times$ speedups on larger problems.
There were more limited speedups on standard user runs (3.6$\times$) where problems are collision dominated and many particles do not even leave a cell.
Additionally, speedup above 8$\times$ is of limited utility without also porting the host hydrodynamic codes to the GPU.

Going forward, the Jayenne development team will work on porting initialization routines (so more time can be spent in transport kernels where GPU speedup can be found), simplifying required integral computations, and porting more features from the CPU version of the code to GPU.

\subsection{MCATK GPU Prototype Code \secsubtitle{Tim Burke, Los Alamos National Laboratory}}

Monte Carlo Application Toolkit (MCATK) is a Monte Carlo particle transport code from Los Alamos National Lab \cite{adams_monte_2015}. 
The current GPU strategy involves moving the ray-tracing component of transport onto the GPU through a package called MonteRay~\cite{sweezy_monteray_2018}.
Ray tracing operations are batched then executed in unison on the GPU.
In radiography applications (where there is a significant amount of streaming) they report an 80$\times$ performance increase.
In simulations with few detectors they see a 1.5--2$\times$ increase in performance.
MonteRay has single precision capabilities.

A prototype Monte Carlo particle transport code is built from stand-alone feature libraries that can be linked together.
Some of these libraries may eventually be open-sourced.
The initial port uses a history based algorithm with both multi-group and continuous energy nuclear data and they see a 10--20$\times$ performance increase on GPUs (Nvidia Tesla V100 vs.\ Intel Xeon Broadwell nodes).
MCATK developers note that implementing an event-based algorithm may improve this further.
In the GPU port, recursion and virtual functions are eliminated. It implements C++ standards 14 and 17 in GPU kernels.

\subsection{LLNL Monte Carlo Transport GPU Status
\secsubtitle{Scott McKinley \& Shawn Dawson, Lawrence Livermore National Laboratory}}


Mercury \cite{mercury_osti_1252640} is a feature rich GPU- and CPU-capable particle transport code for neutrons, photons, and light element charged particles.
Mercury has a sibling thermal radiative transport code, Imp~\cite{imp2019}, which uses an implicit Monte Carlo method.
Imp shares roughly 80\% of its code with Mercury, so advances in hardware implementation provide benefits to both code bases.
The LLNL Monte Carlo Transport Project has spent several years of development optimizing Mercury and Imp to meet the computing potential of new, GPU-centric architectures like Sierra and El Capitan.
The earliest versions of the GPU code struggled to reach desired computational speedup values values~\cite{statusLLNL2019}.
Ongoing optimization efforts have resulted in improved GPU performance~\cite{pozzulp}, although optimization efforts continue
Currently, Mercury and Imp are performant on CPU and Nvidia architectures and are being extended to AMD architectures as El Capitan, an AMD-based machine at LLNL, is being delivered.

This architecture portability is achieved via “loop abstraction”, a process where macros are inserted around parallel loops. The source code in the loop is captured as a lambda expression. This lambda may be launched on the GPU with the standard chevron syntax, or on the host CPUs using OpenMP, or even run sequentially. The macro allows the compiler to generate code for all three scenarios. Where to execute is deferred to run time, and is controllable by the end user.
This method also offers easy iteration and profiling during development, allowing programmers to simply modify a macro when implementing new ideas.
This technique is similar to one used by RAJA~\cite{raja2019}, a portability library developed at LLNL; while the loop abstraction macros may be ported to RAJA in the future, these methods were developed independently.

When porting their code to the GPU, the LLNL Monte Carlo Transport team decided to simply run their existing multiprocessor code on GPUs, use profiling to detect areas of low performance, and then optimize those areas.
This method allowed for some developments that are rare among other transport codes, such as their use of managed memory within the GPU.
Normally, managed memory allocation calls are considered too slow to be performant. 
However, Mercury and Imp use a library called Umpire~\cite{umpire2020} to create a memory pool and amortize the cost of these calls, allowing managed memory to be performant.

Tallying in Mercury and Imp uses atomics.
These atomics can cause memory contention, especially in tallies with low memory usage.
To mitigate this, tally data structures are replicated such that memory contention does not reduce performance.
This often requires full replication on CPUs due to the slower operation of these atomics.

Mercury's history-based particle tracking method on CPUs performed poorly on GPUs.
The code initially used a monolithic kernel that used all registers on the GPU and suffered performance losses due to divergence.
So an event-based algorithm was added.
This was one of many performance improvements that the development team considered, including:
\begin{itemize}
    \item A ``sidecar'' that preserves simulation data between kernel launches,
    \item Single data structure for mesh variables,
    \item Asynchronous loops for independent events, and
    \item Link Time Optimization within Nvidia's compiler (providing 20--35\% speedup on multiple problems).
\end{itemize}
These improvements led to a 5$\times$ speedup on four GPUs compared against 36 CPUs on Sierra.

Development of Mercury and Imp is still ongoing, and there are several challenges the development team is currently facing as they prepare for the MCGIDI, the nuclear data interface used in Mercury is currently being ported to AMD GPUs, and contains many functional structures that are challenging to optimize on GPUs.
LLNL's HPC Toolkit was being used to profile, diagnose, and debug these issues, but the toolkit had problems with MCGIDI, leading to a roadblock in debugging; LLNL's developers are currently addressing these challenges.
The development team has also identified specific areas of difficulty with the port of Mercury and Imp from CUDA to ROCm:
\begin{itemize}
    \item Link Time Optimization takes far longer with ROCm;
    \item The initial port of working CUDA code did not work for all GPU kernels with the ROCm compiler. Some kernels receive segmentation faults within the GPU kernel at runtime which are difficult to diagnose;
    \item NVCC is still used to compile CUDA code, and HIP is only used to compile ROCm code (much like the situation faced by the Shift developers);
    \item Certain optimization techniques in ROCm seem to create segmentation fault issues in Imp, with individual kernels needing to be compiled without optimization; and
\end{itemize}

Overall the Mercury and Imp developers report a successful port to Nvidia GPUs using the NVCC compiler seeing a 5$\times$ node-to-node speedup over 36-core Intel Xeon type nodes. 
They will continue their port to other GPU platforms and continue to try to increase performance of their GPU Monte Carlo codes.

\subsection{High Energy Physics Monte Carlo on GPUs: Celeritas \secsubtitle{Seth Johnson, Oak Ridge National Laboratory}}

Celeritas is a Monte Carlo transport code focused on finding a simulation that matches experimental detector output used in the upgraded High Luminosity Large Hadron Collider (HL-LHC) at CERN \cite{tognini2022celeritas}.
These simulations compute histories that represent analog physical realizations of a particle moving through the detector region of the HL-LHC, referred to as a ``track''.
Currently, these simulations use the Geant4 transport code~\cite{Agostinelli2003} to compute these realizations; however, the massive throughput of data from HL-LHC experiments requires accelerated computing on GPUs.
Celeritas is designed to achieve this by re-structuring the data types and job kernels to run more efficiently on GPUs before passing them to Geant4.
HL-LHC simulations are much more geometry intensive than neutron transport because of the high-energy charged particles being simulated, which requires the inclusion of magnetic fields, continuous slowing down, and multiple scattering.
The simulations being run for the HL-LHC also require a higher degree of accuracy, at scales from millimeters to tens of meters; these problems combine to make these simulations a challenge to implement on GPU architectures.

Celeritas achieves this through a number of new implementations.
Data structures are allocated in chunks on the CPU and passed to the GPU.
This allows the GPU compiler to make better memory optimizations for the problem before running the main particle transport loop.
This loop is written to handle large batches of tracks with an arbitrary number of kernels, using masking to hide unnecessary kernels on tracks where they are not utilized.
This loop runs faster on a per-kernel basis, but sorting overhead reduces this speedup as the number of tracks increases.
This also limits Celeritas to only tracking one set of particles at a time, an ``event-granular parallelism'' that can conflict with the track-granular parallelism expected by Geant4.
The complex geometries used in these simulations cause Celeritas to run with far fewer particles on GPU due to implementation issues the VecGeom
library~\cite{Wenzel2020}. The Oak Ridge Advanced Nested Geometry Engine (ORANGE) has also been considered to alleviate the geometry bottle neck on GPUs.

Despite these issues, Celeritas shows performant speedup when running on GPUs compared to CPUs.
On a range of benchmark problems, Celeritas runs 30--40$\times$ faster on GPUs than CPUs; these results were obtained with both ORANGE and VecGeom, although using ORANGE delivered higher performance metrics.
The GPU saturates during the beginning of these simulations due to electromagnetic showering increasing the number of particle tracks, but these saturated steps of the transport loop account for less than 20 percent of the overall simulation.
Celeritas also shows promise running on CPUs, simulating a detector problem about 20 percent faster than pure Geant4.
Electromagnetic physics problems are still difficult to run on GPUs, but show a roughly 2$\times$ speedup compared to CPU runs.
With problem geometry still proving to be a bottleneck on these problems, Oak Ridge National Laboratory and CERN are both working on developing methods for GPU architectures, which will allow these charged geometries to be simulated more quickly.
GPU performance is also currently limited by the amount of code running on the GPU as opposed to the host CPU; improving data structures and problem geometry for GPUs will likely lead to larger speedups as a result.

\subsection{MC/DC+Harmonize: JIT Compilation \& Asynchronous GPU scheduling \secsubtitle{Ilham Variansyah \& Braxton Cuneo, Center for Exascale Monte Carlo Neutron Transport}}

Monte Carlo Dynamic Code (MC/DC) is built and maintained by researchers associated with the Center for Exascale Monte Carlo Neutron Transport \cite{variansyah_mc23_mcdc}.
MC/DC is fully open source (under a BSD-3 clause license) and written completely in Python with a compilation structure to support highly performant execution on CPU- and GPU-based architectures.
It is written specifically to enable rapid deployment and testing of novel time-dependent transport algorithms and variance reduction techniques on heterogeneous architectures by a diverse user base in an academic setting. 

While development is ongoing and some functionality currently has only experimental support, MC/DC has the following features:
\begin{itemize}
    \item Multigroup physics, including capture, isotropic scattering, and fission (prompt and delayed);
    \item Continuous energy physics (at room temperature via point-wise data generated from NJOY);
    \item Geometry tracking via surface-tracking, quadric CSG surfaces, multi-level lattices, and time-dependent planar surfaces;
    \item Fixed-source (time-dependent) and $k$-eigenvalue simulation modes;
    \item CPU parallel support with MPI;
    \item GPU parallel support with Numba-CUDA (via Harmonize);
    \item Domain decomposition; and
    \item Total reproducibility on any architecture in any simulation mode via hash-based Random Number Generator (RNG) seeding.
\end{itemize}

MC/DC couples a fully transient neutron transport algorithm to a novel (for the field) development scheme.
In particle transport the physics kernels are compute kernels.
This is different from other simulation types that can often rely on exterior packages to offload most of the compute complexity and handle optimizations for heterogeneous architectures (e.g., PetSC \cite{petsc_user_ref}, MFEM~\cite{anderson_mfem_2021}).
Thus, development of particle transport Monte Carlo codes has traditionally involved compiled languages due to speed limitations and requirements of heterogeneous architectures and their APIs.
This can often make development difficult for users outside of a specific lab or field, or those developing on completely different systems, even if theoretically the same architecture.
MC/DC remedies this by using an acceleration and abstraction scheme based on the Numba compiler for Python that supports compilation to CPU and GPU hardware targets~\cite{lam_numba_2015}.
At runtime these compute kernels are lowered into intermediate representations in the LLVM compiler library and compiled to a specific hardware target.
These compute kernels are then bound and executed by a larger Python script.
The MC/DC team studied other acceleration and abstraction techniques (e.g., PyKokkos~\cite{AlAwarETAL21PyKokkos}, PyOpenCL, and PyCUDA~\cite{kloeckner_pycuda_2012}) but determined that Numba was the least intrusive and easiest to use, while still providing significant speed-up for Monte Carlo applications~\cite{morgan2022}.

MC/DC has a full unit, regression, and verification test suite built in and continuously integrated using GitHub Actions.
One verification test is a time-dependent Kobayashi dog-leg benchmark problem~\cite{Kobayashi2001}.
It uses the seven-group C5G7 cross section data~\cite{hou2017oecd} and a dog-leg duct filled with water.
The initial condition is an isotropic pulse of particles in the highest-energy group.
A fissionable region is located in the duct. 
The thermal flux is tallied in three dimensional space and time.
In a strong-scaling analysis, the MC/DC team found that, after a constant 43 second compilation time, the Numba compiler provides a 462$\times$ speedup over pure Python mode.

MC/DC is written in pure Python so that the code is extensible and easily interpreted to users, numerical methods developers, and most maintainers.
When run in pure Python mode (i.e., no acceleration) MC/DC's kernels can be developed and implemented as in any other Python package.
The Numba compiler can be turned on if greater performance is required to run various transport algorithms directly on CPU with compiled binaries that approach compiled code speeds.
While this does restrict use of some Python features (e.g., dictionaries, dynamic typing, support for all libraries) the code is still being developed entirely in Python---just a smaller subset of the language.
When targeting for GPUs, Python-based functions are compiled to an intermediate representation that can be passed to the Harmonize compute framework for further compilation and execution.

Harmonize~\cite{brax2023} is an asynchronous GPU scheduler that can take divergent code bases and dynamically reschedule when kernels get executed such that they binned together. 
It effectively becomes a wrapper around compute kernels to execute them in a way that decreases thread divergence on streaming architectures, thus increasing performance. 
From the Monte Carlo transport perspective this can be thought of as on-the-fly event-based scheduling; however, Harmonize can be generally implemented by any GPU program.

Using these acceleration and abstraction techniques, MC/DC has been able to go from nothing to a fully fledged transient continuous energy transport code with completely novel methods written by relatively novice developers in about two years.
MC/DC developers have done successful explorations into quasi-Monte Carlo \cite{mcdc:variansyah_physor22_pct}, hybrid iterative techniques for $k$-eigenvalue simulations \cite{mcdc:qmc, mcdc:qmcabs, mcdc:pasmann_mitigating_2024}, transient population control techniques \cite{mcdc:variansyah_nse22_pct}, automatic transient weight window production, hash-based random number generation, global uncertainty quantification \cite{mcdc:clements_mc23}, residual Monte Carlo methods, and machine learning techniques for dynamic node scheduling---among other active areas of transient methods exploration.
As the benefit of some of these novel algorithms is proven in MC/DC, developers look to implement in other software packages, further pushing transient radiation transport methods into the future.

\subsection{OpenMC at Exascale: Performance Portable Monte Carlo Particle Transport on Intel, Nvidia, and AMD GPUs \secsubtitle{John Tramm, Argonne National Laboratory}} 
\label{sec:openmc}

OpenMC is a Monte Carlo full-featured neutron and photon transport code \cite{RomanoPaulK.2013TOMC}. 
For years, OpenMP threading has served as OpenMC's
shared-memory parallelism model, with MPI available for distributed memory
parallelism. 
\cite{trammPortableGPUAcceleration2022} recently refactored a branch of OpenMC to utilize OpenMP5 target offloading technology.
This interface allows for shared-memory multiprocessing on multiple types of CPU and GPU platforms; as such, it is considered a ``portable'' interface that does not require modification to run on its supported architectures.
This portability was the main motivation for the development of OpenMC, as there are many different architectures in use across the U.S. national laboratories, particularly in the realm of GPU programming.
Device offloading was a new feature in OpenMP when development for OpenMC began; this was noted as a considerable risk for the development team, as offloading is used extensively in Monte Carlo codes.
It was mentioned in discussion that this risk led to disappointing performance levels of several other OpenMP based (non-Monte Carlo) codes, and that OpenMC was fortunate enough to have a direct line of contact to OpenMP developers to help debug issues that arose during development, which aided in overcoming this risk.

In a depleted pin-cell benchmark study OpenMC on CPU performs about the same as other Monte Carlo transport codes on dual socket Intel Xeon Platinum 8260 nodes with 48 cores in total.
In that same study on GPU the OpenMC development team reports speedups (as compared to a CPU runtime) of 5.4$\times$ on a single Nvidia A100 GPU, 7.0$\times$ on a single AMD MI250X GPU, and a 12.8$\times$ on a single Intel Ponte Vecchio GPU.
Intel GPUs use the OpenMP compiler, which is supported by internal portability codes such as Kokkos and RAJA.
RAJA developers mentioned in discussion that, as long as the compiler is mature enough to be publicly deployed, implementations of the same code using SYCL, HIP, and OpenMP backends have shown to be equivalently performant.

These performance metrics were achieved through a number of optimization methods implemented by the development team. 
Thankfully, very little specialized optimization was required to port the pre-existing Monte Carlo code into OpenMP, aside from some by-device optimizations for sorting particles. 
The major areas of optimization mentioned in this presentation were tallies and cross-section lookups.

Cross-section lookups were by far the most expensive kernel due to non-coherent memory access. The cost of this kernel was reduced with two optimizations. The first was to sort particles before executing the kernel. 
This sorted lookup results in faster performance because these loaded values are often shared between threads in a single warp. 
The second was to remove the microscopic cross section cache and perform extra lookups on GPU. This was faster due to the reduction of data required to move between the the CPU and GPU as well as on board high bandwidth GPU memory.
However, the microscopic cross-section caching method is still more performant on CPUs, and is still implemented on those architectures.
Additionally, the development team found that launching multiple MPI ranks per GPU seemed to saturate the GPU more quickly, leading to significant speed increases; in particular, GPU performance on Intel GPUs is roughly double that of AMD or Nvidia GPUs. 

During the development of OpenMC, some computing trends were noted that could inform future development. High-bandwidth memory processors did not seem to improve the performance of OpenMC, and do not seem to be a necessary avenue for future development. It is also worth noting that CPU performance on Monte Carlo simulations seems to double every five years, while GPU performance doubles every two years. When accounting for power consumption, the future seems even more bleak for CPUs; while GPU performance still manages to double every three years, CPU performance is seemingly stagnant. During discussion, it was also mentioned that AI accelerators have shown promising performance boosts; however, they are currently much more difficult to design software for than conventional GPU accelerators, which could lead to trouble as development gets more intricate.

\subsection{Experience Adapting OpenMC to CUDA \secsubtitle{Gavin Ridley, Massachusetts Institute of Technology}}

In parallel to efforts at Argonne National Laboratory to adapt OpenMC for GPU computation,
a smaller thrust at the Massachusetts Institute of Technology focused on a port using
the more widely adopted CUDA toolchain~\cite{ridleyDesignOptimizationGPU2021}.
The software development sought to minimize the source code change between the CPU-based
code and the GPU-based code, potentially incurring a performance cost. 
About 50,000 lines of C++ source code comprise the core operations which must be accelerated by GPU.

This work employed the same event-based tracking technique as the OpenMP-focused work,
and provided simulation results identical to the CPU and OpenMP GPU code within the usual
floating point associativity differences. In contrast, the conceptual modification of the
source code was minimized. Because OpenMC extensively expresses abstraction with polymorphic
data structures, and CUDA can use polymorphism at some hard-to-quantify performance cost,
the code used polymorphism. In addition, a few core STL data structures were replaced with
implementations that can compile as CUDA device code. In particular, we created an implementation of \texttt{std::unique\_ptr} that can maintain separate copies of data
on both the host and device in a way that maintains separate virtual tables on each.
A threadsafe implementation of \texttt{std::vector} was also written to minimize the
required programming effort.

The custom smart pointer implementation was necessitated by the fact that objects exhibiting polymorphism must be constructed
on the GPU device. The nuclear data readers in OpenMC run on CPU, so the code copied the
nuclear data structures to device first, then executed move constructors called in-place
to set up device virtual tables.

In contrast, because the OpenMP port could not support polymorphic design patterns,
the source code required substantial reworking. All polymorphic behavior had to be
replaced with switch-case statements: an error-prone approach which led to at least
one persistent and challenging bug. The OpenMP port also requires extensive manual memory management,
whereas the MIT version employed CUDA unified memory to automatically manage memory.
At the same time, because the nuclear data memory was not manually distributed, the
CUDA code likely exhibited a substantial performance degradation due to a higher
frequency of uncoalesced memory accesses. This confounds quantification of the performance degradation experienced due to higher software abstraction in this approach.

We also experimented with a structure-of-array approach rather than array-of-structures
for particle data, but found only modest performance gains compared to the array-of-structure approach, around 5\% faster calculation rates at best.

After applying the cache-less microscopic cross section trick mentioned in the prior 
OpenMC section and by  \cite{trammPortableGPUAcceleration2022}, along with a variety of
other optimizations such as trading occupancy for reduced thread divergence, the CUDA-based
code achieved a performance about a factor of two worse than the OpenMP code.
Given the widespread use of Monte Carlo transport codes, our recommendation for Monte
Carlo particle transport code developers going forward is to use flat data structures without
polymorphism. Since optimal performance is important here,
the extra programmer effort required to avoid software abstraction pays off in the
long run.

\subsection{Monte Carlo Research at NNL \secsubtitle{Paul Burke, Naval Nuclear Laboratory}}

MC21 is the Naval Nuclear Laboratory's primary Monte Carlo code. It is used internally for reactor physics, criticality safety, and shielding analysis and is written primarily as a production design tool \cite{GriesheimerD.P.2015Mv–A}. 
MC21 has support for:
\begin{itemize}
    \item Continuous-energy transport;
    \item Fixed source and eigenvalue calculations;
    \item Neutron and photon transport capabilities, including coupled transport;
    \item Flexible constructive solid geometry system with support for hierarchical model construction;
    \item In-line feedback effects (depletion, T/H, Xenon, eigenvalue search (with movable geometry), neutron and photon heating);
    \item Generalized tally capability (via user-defined “phase filters” to control which particles are allowed to score each tally);
    \item In-line decay source; and
    \item Weight-window variance reduction (with support for coupled radiation simulations).
\end{itemize}
These features allow MC21 to model problems as geometrically complex as the advanced test reactor with full geometry and physics. 

GPU support in MC21 is only available in a developmental fork and parallelism schemes are governed by machines available at NNL. MC21 currently uses both OpenMP and MPI parallelism, primarily targeting Intel Xeon-type machines. Using this tact, simulations have been performed at the peta-scale. When approaching GPU porting, the MC21 development team has focused on minimizing code touched by a port and maintaining compatibility with in-line feedback, searches, and tally systems. The team started by selectively porting only bottle-necking kernels beginning with macroscopic cross section lookups for tracking~\cite{akmc21}. Tracking and tallying is done on the CPU, then when a macroscopic cross section is needed that particle is banked and tracking is paused. This continues until a sufficient number of cross section lookup calls is batched then they are executed all at once on the GPU. 
While mini-app results were promising, when implemented in the full code only a 10\% \textit{speedup} in the most-lookup-bound case was found with a 30\% \textit{slowdown} for more representative simulations.

Other GPU developments have primarily involved ray tracing capabilities both for MC21 simulations and for visualizations in NNL's internal visualizer MCVIZ.
The two code stacks use the same kernel, code-named Rayzor.
Again here the desire is to limit code alterations required, and enable compatibility with the full feature-rich code bases. 
This is different from other development teams which often allow for divergent code bases and different features for different architecture targets.

In the ray tracing algorithm, Rayzor replaces recursive and polymorphic call stacks with nested for loops via a disjunctive normal form expression, decreasing thread divergence. Any arbitrary expression can be converted into disjunctive normal form and typically model components are very close to this form already.
Also Rayzor will use a bounding volume hierarchy scheme as opposed to a constructive solid geometry representation.
This is a commonly used algorithm for ray tracing in which the search space is divided in two branches at each level of the tree, with known bounding boxes.
This is different from constructive solid geometry as tracking continues through bodies and does not stop at the surface.
It also requires component hierarchy where each component with children needs its own bounding volume hierarchy.
Generally they have identified that a GPU port requires removal of polymorphism and recursions, a move to \textit{structures of arrays} for data storage restructuring of compilation units.

Going forward the MC21 development team has identified a primary performance  milestone for significant user workflow improvement as two orders of magnitude of speedup.
Speedups of lesser magnitude represent a quality-of-life improvement but do not disrupt the current workflow dynamics of diffusion analyses during the initial design phase with Monte Carlo results only needed when moving towards production.
Developers from the NNSA laboratories (LANL, LLNL, and SNL) discussed a similar paradigm with their users only willing to upgrade to a new version of a particular software if there was at least a 10$\times$ speedup, with a 100$\times$ speedup to change actual the design process.

The MC21 development team will continue to approach a GPU port by minimizing the amount of code touched and maturating full feature compatibility.
This is due to the aforementioned nature of the MC21 use-case (being for reactor design) and the high cost associated with re-verification of models, in addition to novel methods.
The primary areas of development include focusing on hybrid acceleration schemes and visualization/mesh homogenization while continuing down a piecemeal-port for GPU offloading, with the primary goal to get users accelerated functionality through smaller, single-use applications.

\section{Standardization of Test Problems}
Although these codes are largely designed to solve similar types of problems, the various benchmarks and challenge problems used to report performance make it difficult for users, code designers, and others to reach a consensus when comparing the performance of different codes. 
For this reason, a discussion was facilitated for standardizing these test problems such that their results could be more easily compared across architectures, code bases, and HPC systems.

Several figures of merit were discussed, including segments per second and particles per second.
It was noted that particles per second could work as a standard figure of merit if all codes solved the same benchmark problem with the same features, whereas segments per second could be more easily compared across disparate benchmarks. 
However, each figure of merit comes with its own problems; namely, segments per second is difficult to track, especially when additional features like variance reduction and implicit capture are added, while particles per second is not as universal as it seems, with particle weights and splitting handled differently between code bases. 
Adding a wide range of figures of merit may be the best compromise available, giving different code bases with different features a number of potential points of comparison to use.

Along with deciding on one or more figures of merit, summit members wished to find, or build, a benchmark problem that could be simulated using as many code bases as possible. 
It quickly became clear that this idea would be difficult due to the different specializations each of these codes pursued---benchmarks that work well on OpenMC may not work at all on Celeritas, for example. 
As a result, an idea was floated on the other end of the spectrum: a benchmark would instead be created for each type of problem the collective code bases were designed to solve.
Several existing repositories for collecting problems and runtime results were posited as a candidate for this idea, including an open-source repository being developed by the American Nuclear Society's Mathematics and Computations Division, and a catalog maintained by ICSBP, although the latter was deemed too complex for use by most codes.

The majority of the challenge problems discussed were neutron transport problems.
Some discussion arose around whether the summit members wanted a \textit{benchmark problem}, which measures the correctness of a code base, or a \textit{challenge problem}, which tests the performance of a code base at the limits of its operational capacity.
Developers from NNSA labs recommended a Godiva sphere criticality test while those from Oak Ridge proposed a continuous energy reactor transport problem.

By the end of the discussion, the majority of summit members determined a time-dependent continuous-energy form of the C5G7 challenge problem~\cite{hou2017oecd} to be a suitable candidate for a standardized test problem.
A unified figure of merit was identified: average reaction tallies across each pin region of the problem; the problem would be modified by developers at Oak Ridge National Laboratory to allow for this figure of merit to be reported.
Additional features, such as multi-group and implicit capture, are also to be added as toggle-able options to the base problem; however, the base problem was chosen as the standard for all Monte Carlo codes for the reporting of comparable figures of merit.

\section{Outlook}

The major takeaways from this summit were concerned with collaboration.
It is clear from the efforts made by the lab members present that GPU development still represents a formidable but necessary challenge, especially with upcoming AMD and Intel architectures present in the exascale-class supercomputers of the future.
The use of multiple co-functional code bases among the national labs and academia remains a useful tool for comparing problem-solving methodologies, as well as verifying and validating results.
To aid in these comparisons, a more standard set of generalized benchmark problems should be produced and distributed with the performance results open and available.
Additionally, a discussion arose around the future of Monte Carlo and its place in the national labs, as well as public and private industry development.
Those present at the summit agreed that the Monte Carlo development community, and perhaps the neutronics community at large, should develop a report projecting how these codes can continue to develop in the near future, and how best to modernize and contextualize Monte Carlo as a useful tool in modern industry.

Moreover, the environment created by the summit was noted by all attendees to be productive and refreshing.
Multiple summit members voiced a desire for more events like this in the future, where developers can share their advancements and insights through presentations and discussion rather than publications.
Attendees were also glad to receive immediate feedback from one another, and there was a good deal of useful ideation and deliberation over both days of the summit.
Overall, both the organizers and guests came away from the event feeling it was a great success.
These successes have led members of CEMeNT to consider making the Monte Carlo Computational Summit a recurring event and expanding to include other developers outside of the Center and national labs.

\section{Acknowledgments}

We would like to thank the attendees, presenters, and organizers including co-chairs Alex Long and Steven Hamilton. We would also like to thank the University of Notre Dame and ND Energy for hosting the Monte Carlo computational summit. Joanna Piper Morgan has previously had a co-op position at Advanced Micro Devices (AMD) and previosuly interned at Los Alamos National Laboratory where she worked on MCATK. All other authors declare no conflicts of interest.

\vspace{0.1in}
This work was supported by the Center for Exascale Monte-Carlo Neutron Transport (CEMeNT) a PSAAP-III project funded by the Department of Energy, grant number: DE-NA003967.

\bibliographystyle{IEEEtran}
\bibliography{refs}
\newpage

\end{document}